\documentclass[preprint]{iucr}
\usepackage{graphicx}
\usepackage[utf8]{inputenc}
\usepackage{url}
\usepackage{srcltx}             

\usepackage{dcolumn}       
\newcolumntype{d}[1]{D{.}{\cdot}{#1}}
\newcolumntype{.}{D{.}{.}{-1}}
\newcolumntype{,}{D{,}{,}{2}}

\def\nba#1{}                  

     \paperprodcode{a000000}      
     \paperref{xx9999}            
     \papertype{FA}               

     \paperlang{english}          
     \journalcode{J}              
     \journalyr{2010}
     \journaliss{1}
     \journalvol{42}
     \journalfirstpage{000}
     \journallastpage{000}
     \journalreceived{0 XXXXXXX 0000}
     \journalaccepted{0 XXXXXXX 0000}
     \journalonline{0 XXXXXXX 0000}

\begin{document}                  



\title{SrRietveld: A program for automating Rietveld refinements for high throughput powder diffraction studies}


\author[a,b]{P.}{Tian}
\author[b]{W.}{Zhou}
\author[b]{J. }{Liu}
\author[b]{Y.}{Shang}
\author[b]{C.~L.}{Farrow}
\author[b]{P. }{Juh\'as}
\cauthor[b,c]{S.~J.~L.}{Billinge}{sb2896@columbia.edu}

\aff[a]{{Department of Physics and Astronomy, Michigan State University}
\city{{East Lansing}, Michigan, 48824, \country{USA}}}
\aff[b]{{Department of Applied Physics and Applied Mathematics, Columbia University}
\city{{New York}, New York, 10027, \country{USA}}}
\aff[c]{{Condensed Matter Physics and Materials Science Department, Brookhaven National Laboratory,}
\city{{Upton}, New York, 11973, \country{USA}}}

\maketitle                        

\begin{synopsis}
SrRietveld, a Python based Rietveld refinement program is described.
SrRietveld extends and automates the popular existing Rietveld programs
FullProf and GSAS to facilitate studies that involve many refinements
such as the analysis of large quantities of data-sets.
\end{synopsis}

\begin{abstract}
SrRietveld is a highly automated software toolkit for Rietveld refinement.
Compared to traditional refinement programs, it is more efficient to use
and easier to learn. It is designed for modern high throughput
diffractometers and capable of processing large numbers of data-sets
with minimal effort.  The new software currently uses conventional Rietveld
refinement engines, automating GSAS and FullProf refinements.
However, as well as automating and extending many tasks
associated with these programs, it is designed in a flexible and extensible way
so that in the future these engines can be replaced with new refinement engines as they
 become available. SrRietveld is an open source
software package developed in Python.
\end{abstract}


\section{Introduction}
\label{Introduction}


The Rietveld refinement method is widely used for determining crystalline material
structures from x-ray or neutron diffraction data~\cite{rietv;jac69,rietv;ac67}.
In this method, a theoretical line profile is calculated from a structure model that is
refined using a least-squares approach until it matches the discrete data from
neutron or x-ray powder diffractometers. Many computer programs have been
developed for Rietveld
refinement~\cite{larso;unpub04,rodri;pb93,murra;pdpl90,howar;lhrl98,izumi;trj89,bergm;cpd98,coelh;unpub04}.
However, most software programs of this type require intensive user inputs and
there is also a sharp learning curve for new users. The high data throughput
from new generation
diffractometers~\cite{proff;apa01i,willi;pb98,mason;pb06,hodge;sns09},
such as POWGEN at the Spallation Neutron Source (SNS) at Oak Ridge National
Laboratory, is more than the conventional Rietveld refinement
software can easily handle. Poor convergence of current programs is another
obstacle for accurate determination of refined parameters and conventional
refinement software often diverges, requiring significant human intervention to
find the optimal structure solution. SrRietveld is designed to improve the user experience in
computer aided Rietveld refinement.

The new software makes use of the existing refinement
programs GSAS~\cite{larso;unpub04} and FullProf~\cite{rodri;pb93}, which are
already widely used. These programs are used by SrRietveld as refinement
engines and SrRietveld provides an automation layer and a graphical user interface.
The new functionalities are implemented in Python packages that can
manipulate and communicate with the refinement engines. The refinement process
can be controlled from SrRietveld and the results can be displayed and analyzed.
This architecture separates the
controlling scripts from the refinement engines and also allows the
implementation of other engines in the future. Using SrRietveld, average
users can set up high-throughput refinement quickly and save time in analyzing large
numbers of data-sets. Meanwhile, advanced users are still able to customize
the refinement routines, create their own Python scripts, or even develop new
software applications based on SrRietveld's functionalities.  The SrRietveld
codes are open source.

\section{SrRietveld}

\subsection{Design Principles}

SrRietveld has been developed as part of the NSF-funded distributed data
analysis for neutron scattering measurements (DANSE) project in the Python
language~\cite{python;web} using object oriented programming (OOP) concepts.
Python
is extensively used in scientific software development. It is cross-platform
and suitable for fast development. Also there are many well designed packages
for scientific programming and visualization, such as
Matplotlib~\cite{matplotlib;web}, NumPy~\cite{numpy;web} and
SciPy~\cite{scipy;web}. The OOP design enables SrRietveld to be easily
maintained and extended.

Refinement information and results are
stored in a flexible HDF5 data structure~\cite{hdf5;web} allowing easy organization of a large
number of refinements.  SrRietveld also provides powerful tools for visualizing refinements.  Fits
can be plotted in real-time during refinement, as can selections of refined parameters
plotted against refinement number or some other meta-parameter such as
temperature.
Pre-written scripts are provided for automating common tasks, such as refining
from coarse guesses of initial parameters, sequential refinement, etc.
It is also possible for advanced users to develop their own macro scripts
and contribute to the development of SrRietveld. Users are supported through
an online community where features can be requested and bugs reported.

\subsection{Implementation}



\begin{figure}
   \begin{center}
     \includegraphics[clip=true,width=0.9\textwidth]{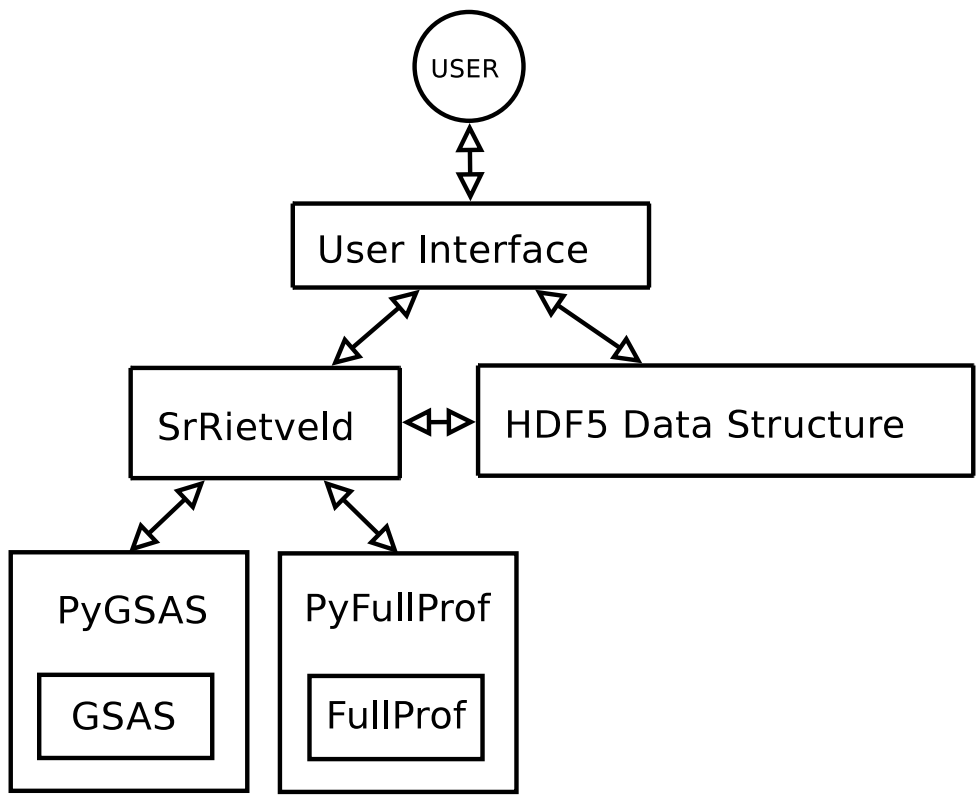}
     \label{fig;srrietveld_structure}
     \caption{SrRietveld software architecture. The User Interface, SrRietveld, HDF5 Data Structure, PyGSAS, and PyFullprof are python packages in the SrRietveld project. The GSAS and Fullprof represent the underlying refinement engines.}
   \end{center}
\end{figure}

The architecture of SrRietveld is shown in Figure~\ref{fig;srrietveld_structure}.
SrRietveld consists of several programmatic units, indicated by boxes in Figure~\ref{fig;srrietveld_structure}.  The SrRietveld user
interface is a full-featured graphical user interface (GUI) that enables the
user to conveniently configure the refinement routines, modify inputs, and
investigate and analyze the results.  The GUI interacts with a control layer
that coordinates communication with the refinement engines, PyGSAS and
PyFullProf. PyGSAS and PyFullProf are Python libraries that give programmatic
access to GSAS and FullProf.  The implementation of PyFullProf, PyGSAS and the control
layer ensures that SrRietveld has consistent behavior even while using
different engines, allowing refinements to be readily carried out on the same
data using either engine. The HDF5 data structure has been designed to be relatively small in size and quick to load, even when handling thousands of datasets.

SrRietveld incorporates most of the major functionality supported by the
current versions of FullProf and GSAS. However, some features, such as the
refinement of magnetic phases, are not implemented, though this is
planned for a later release.

\begin{figure}
\begin{center}
   \includegraphics[clip=true,width=1.0\textwidth]{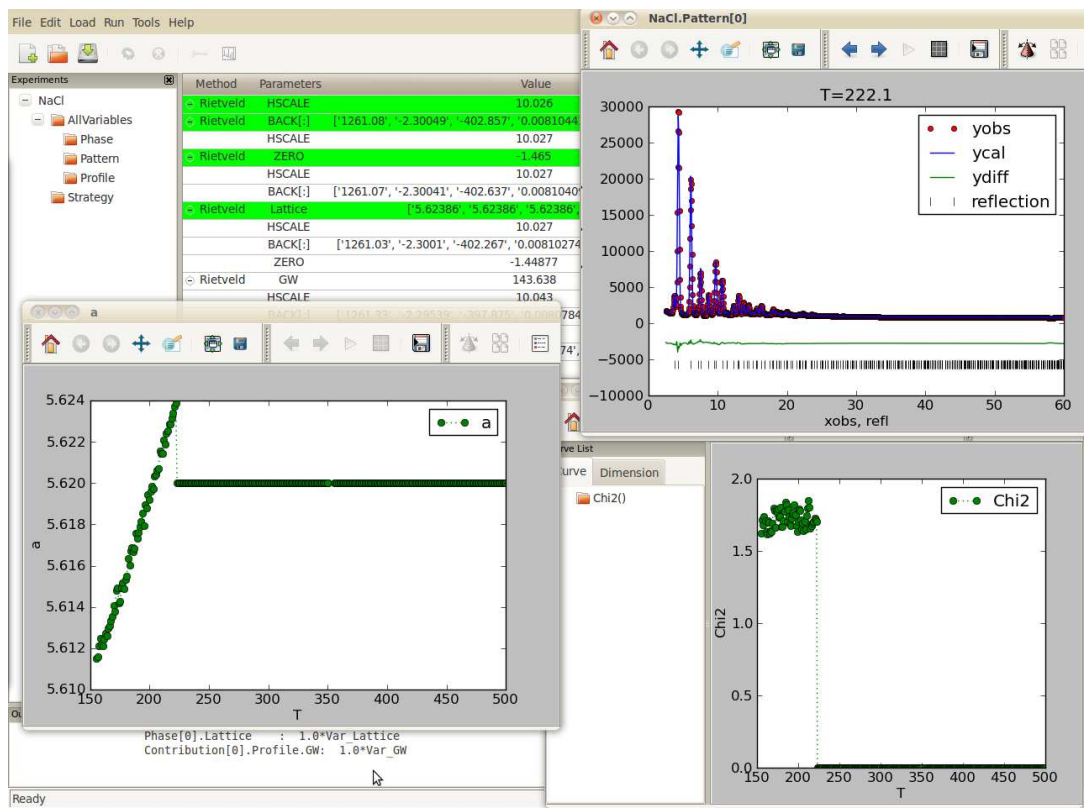}
	\caption{Screen-shot of SrRietveld's plotting window during a
	sequential refinement. The data are NaCl which were measured as a
	function of temperature (see Section~\ref{sec;eg} for details).
	The top right panel shows the fit of the dataset being currently
	refined. The crosses are the data, the solid red line is the
	calculated pattern and the green-line offset below is the difference
	curve. The black tick-marks show the positions of Bragg peak reflections in the model.
    We have chosen to monitor the $a$-lattice parameter values (bottom left panel) refined
	from each of the data-sets. The $\chi^2$ of the
    fits are shown in the lower right panel, allowing issues to be addressed as they emerge.
    Text outputs from the program are displayed in a text window that is just evident in the lower left of
    the Figure.  \label{fig;screenshot}}

\end{center}
    \end{figure}

The program captures errors and attempts to recover automatically when a
refinement diverges, and thus increases the robustness of this sequential
refinement process.
Figure~\ref{fig;screenshot} shows a screen-shot of SrRietveld when
it is carrying out sequential refinements on NaCl x-ray powder diffraction
data collected as a function of temperature, as described in detail in the Figure caption.
In addition to the ease-of-use, SrRietveld provides flexibility and
extensibility to advanced users. Since SrRietveld is modular and designed with
focus on extensibility, the default behavior of the software can be redefined.
Advanced user scripts can be written in Python to give new functionalities.
Such a script is demonstrated below, which fits a Debye model for the lattice
dynamics to the temperature dependence of the refined thermal parameters. In
future releases we will allow easy incorporation of user-defined scripts into the
program for easy sharing.
Scientific programmers can also implement the application programming
interfaces provided in SrRietveld into their own applications so SrRietveld functionality can be incorporated into other programs.

SrRietveld is open source software distributed under the BSD
License~\cite{bsd;web}. It is free to use, subject to the copyright
restrictions and disclaimer, though we ask that papers published from
work done using SrRietveld cite this paper, as well as the paper describing
the particular refinement engine used (FullProf or GSAS).
More information can be obtained from the project
web pages~\cite{srrietveld;web} or by contacting professor Simon Billinge (sb2896@columbia.edu).

\subsection{Future Developments}

SrRietveld is under active development and we encourage users to post bug reports and feature requests
on the SrRietveld online community group (details in the documentation).
The plans for future releases include
support for parallel computation, magnetic diffraction,
and features for engineering diffraction analysis.

\section{Example}
\label{sec;eg}

To demonstrate a typical use case, sequential refinements were carried out on
x-ray powder diffraction patterns measured from NaCl at a series of temperatures.
The data were collected at beamline 11-ID-B of the Advanced Photon
Source (APS) at Argonne National Laboratory (ANL). The NaCl powder was
obtained from
Puratronic\textregistered{} with 99.999\%, 5N grade purity. To demonstrate the
ability to handle data from a high throughput instrument, many data-sets were
collected using a low-resolution high throughput mode~\cite{chupa;jac03}. The sample was contained
in a 1~mm diameter kapton tube mounted perpendicular to the synchrotron beam and
cooled using an Oxford cryostream liquid nitrogen cooler. Data were collected
at 307 temperature points between 155~K and 500~K on a 2D Perkin-Elmer
amorphous silicon detector mounted 128~mm behind the sample and perpendicular
to the beam. The beam energy was 58.26~keV giving an x-ray wavelength of
0.2128~\AA. The temperature was ramped continuously at 3~K/min during data collection.

To improve the convergence during the sequential refinement, the refined
results from the previous temperature point were used as initial values in the
refinement on the data set at the following temperature point. The first point to
be refined was the lowest temperature point. A refinement scheme was used whereby
parameters were turned on in turn according to normal
practice~\cite{young;b;trm93} beginning with a Lebail refinement~\cite{lebail;mrb87} of the lattice
parameters and the background, followed by Rietveld refinements where the scale parameter,
zero shift parameters, lattice parameter, etc., were switched on in turn. This procedure is automated
in SrRietveld and requires no user input after initial setup.  Although in principle in a sequential
refinement it should not be necessary to follow this scheme at every step since the initial parameters
from the previous refinement are used and are already close to the minimum in practice we find that
at some temperature points the refinement will still diverge if this scheme is not followed.  However,
SrRietveld provides great flexibility.  For example, if speed of refinement is an issue the scheme could be changed so that all parameters are switched on at once after the first refinement, and the step-by-step scheme only turned on if a refinement diverges.  Because SrRietveld controls the engines with scripts, there is great flexibility for automating such a procedure.

\begin{figure}
  \begin{center}
    \includegraphics[clip=true,width=1.0\textwidth]{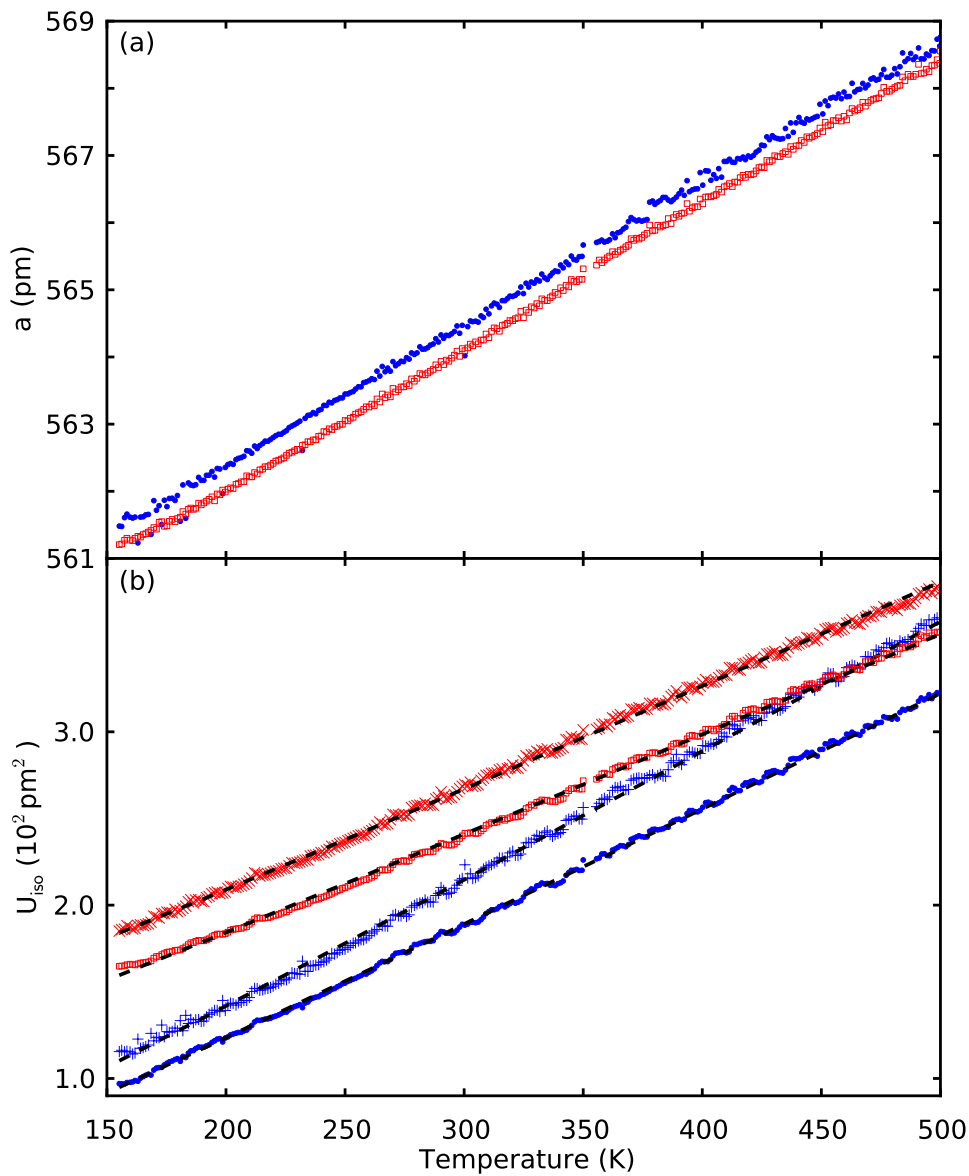}
      \label{fig;bisocl12}
	  \caption{Temperature dependence of SrRietveld refinement
          results for NaCl using FullProf (blue dots) and GSAS (red squares) engines.
          (a) lattice parameter, $a$, and
          (b) atomic displacement parameters, $U_{iso}$, at Na (crosses for FullProf, plus markers for GSAS)
          and Cl (dots for FullProf, squares for GSAS) sites.  Dashed lines mark fitted Debye
          model curves.
          }
  \end{center}
\end{figure}

In the NaCl refinements, the full-set of refinable
parameters used was the scale factor, the zero shift factor, the lattice parameter,
the background parameters, the peak profile parameters, and the isotropic thermal
displacement parameters. The starting model was taken from the
literature~\cite{itable;volc95}.  Refinements were carried out on the same
data using both GSAS and FullProf engines within SrRietveld.  Both temperature
series refinements used the same SrRietveld script, with the refinement engine
changed from FullProf to GSAS by simply loading a different type of
of template in the GUI.
The ease of refining using both GSAS and FullProf engines is
one of the great strengths of SrRietveld as we show here. Although GSAS and
FullProf models are highly similar, there are subtle differences in how they
handle experimental effects. For example, the way they handle the
background function is quite different. In the FullProf example, user defined
points are imported and a linear interpolation is applied between them for background correction.
In the GSAS case, the background function is fitted to the data together
with Rietveld profile.  We have used the background function 6,
a power series with negative power terms~\cite{larso;unpub04}).  Also, the peak profiles are slightly
different. The pseudo-Voigt peak profile function was used in the FullProf case.
The GSAS refinements used profile function~3, which is also
a pseudo-Voigt, but parametrized in a slightly different way~\cite{larso;unpub04}.  These model differences result in
slight differences in the physical parameters refined from the data from each engine, as
evident in Fig.~\ref{fig;bisocl12} and Table~\ref{tab;rtvalues}.
%
%
%
\begin{table}
\label{tab;rtvalues}
\caption{Rietveld refinement results for NaCl at room temperature. }
\begin{tabular}[c]{l...}
\hline\hline
                             & \multicolumn{1}{c}{$a(pm)$} & \multicolumn{1}{c}{$U_{iso, \mathrm{Na}}(pm^2)$} & \multicolumn{1}{c}{$U_{iso, \mathrm{Cl}}(pm^2)$} \\
\hline
GSAS                         & 564.00(5)                   & 267.2(3)                                         & 238.46(13) \\
FullProf                     & 564.35(9)                   & 214.4(6)                                         & 187.3(3) \\
literature*                  & 563.9^\dag                  & 22.9(4)                                          & 189(4) \\
\hline
\multicolumn{4}{l}{
* \cite{butt;ac73}
$^\dag$ measured at 295~K
} \\
\hline\hline
\end{tabular}
\end{table}
%

Another feature of SrRietveld is that the parameters from many refinements are
retained in the Python data-structure, allowing post-processing scripts of
arbitrary complexity to be written. Here this feature is illustrated by writing a script
to fit the Debye model to the refined Debye-Waller factors from Na and Cl from
each of the engines. It is possible for advanced users to make plug-ins for
SrRietveld, which can be saved and reused by the user, and also readily shared
with other users.

The theoretical values of $U_{iso}$ are calculated from~\cite{fisch;jpc78}
\begin{equation}
  U_{iso}\left( T \right) = \frac{1}{8\pi^2}\left\{ \frac{6h^2}{Mk_B\theta _D}\left[ \frac{1}{4} + \left( \frac{T}{\theta_D}\right)^2\int_{0}^{\theta_D/T}\frac{xdx}{e^x-1}\right] \right\} + \sigma_o
\label{eq;debye}
\end{equation}
where the refinable parameters are $\theta _D$ and $\sigma_o$, the Debye
temperature and offset, respectively. Additionally, $T$ is temperature in Kelvin,
$M$ is the mass of the ion in question, $k_B$ and $h$ are the Boltzmann and
Planck constants, respectively. The fitting to the refined $U_{iso}$ data is carried
out with the least square optimization algorithm available in the
SciPy Python scientific computing package~\cite{scipy;web}. The best-fit lines are shown in
Fig.~\ref{fig;bisocl12} as dashed lines through the data and the
quantitative results from Debye fitting are listed in Table~\ref{table;biso},
where $\Theta_{D, Na}$ and $\Theta_{D, Cl}$ represent the Debye temperatures
 from the fittings on $U_{iso}$ values of Na and Cl, respectively.
%
\begin{table}
\caption{Debye temperatures of NaCl from fitting the Debye model to
$U_{iso}\left( T \right)$ values refined using different
refinement engines.  The numbers in brackets are the estimated standard deviation on the last digit.\label{table;biso}}
\centering
\begin{tabular}{ l....}
\hline
\hline
         & \multicolumn{1}{c}{$\Theta_{D,\mathrm{Na}}(K)$} & \multicolumn{1}{c}{$\sigma_{o,\mathrm{Na}}(pm^2)$}  & \multicolumn{1}{c}{$\Theta_{D,\mathrm{Cl}}(K)$}  & \multicolumn{1}{c}{$\sigma_{o,\mathrm{Cl}}(pm^2)$} \\
\hline
GSAS     & 322.4(3)                                        & 78.6(3)                                             & 266.6(3)                                         & 61.5(3) \\
FullProf & 289.2(4)                                        & -18.1(5)                                            & 248.5(2)                                         & -16.9(2) \\	
\hline
\end{tabular}
\end{table}
%

The differences between the Debye temperatures and offsets refined from
each engine are much larger than the estimated standard deviations.  The esds give a measure
of the precision of the fit, and not the accuracy of the Debye temperatures.  To understand
this, note that the dashed lines fit well to the $U_{iso}$ curves, but the slopes and offsets
of the curves are clearly different.  The differences in slopes and offsets come from differences in the mathematical
model for the line-shapes and backgrounds used in FullProf and GSAS, respectively.  This
illustrates the advantage of being able to refine data using two refinement engines.  If the FullProf and GSAS models are equally valid and we cannot say one or the other is inferior, then the dispersion of results from the two engines gives a realistic estimate of our uncertainty in the accuracy of the refined parameters.  In this case we may take into account the result from both engines to get estimate of $\Theta_{D,\mathrm{Na}} = 305\pm 17$~K and $\Theta_{D,\mathrm{Cl}} = 257\pm 9$~K where the estimated uncertainties now reflect accuracy rather than precision.  The corresponding offset parameters obtained by combining the Rietveld and GSAS results are $30 \pm 50$~\AA$^{-1}$ and $20 \pm 40$~\AA$^{-1}$ for Na and Cl, respectively.  Therefore there is no evidence for a static offset in the Debye model that would indicate a significant number of static defects in the structure.

In other experiments, the Debye temperature of NaCl can be determined either
by measuring the elastic constant or by fitting the specific heat data. The
$\Theta_D$ values measured with these two methods are $321.2\pm 1.6$~K and
$320.6\pm 1.5$~K, respectively, reported by Lewis et. al.~\cite{lewis;pr67}.
Broadly speaking the Debye temperatures measured here are in good agreement.
Because of the simplicity of the Debye model, and the different way that
different measurements weight the phonon density of states, the different
approaches are not expected to yield the same Debye temperatures. Furthermore,
the thermodynamic measurements give some average of the full density of states
whereas the diffraction measurement differentiates the behavior of Na and Cl.
The agreement between the values determined from diffraction and those obtained
from other methods are therefore quite acceptable.


\section{Acknowledgments}

The refinement engines used in the current release of SrRietveld are FullProf
and GSAS. Along with many others in the community, we would like to thank Juan
Rodriguez-Carvajal and the other FullProf developers and Robert B. Von Dreele,
Brian H. Toby and the other GSAS developers, for their enormous efforts in
developing FullProf and GSAS. We would also like to thank the other developers
in the DANSE project for useful advice and help. We appreciate the cooperation
and valuable suggestions from Emil Bo\v{z}in, Jason Hodges, Ashfia Huq, Ke An,
Paolo Radaelli and Laurent Chapon. We thank Timur Dykhne for help with the documentation.
In addition to their suggestions, Aziz
Daoud-Aladine, Jon Hanson, Vanessa Peterson, and Andrew Studer shared data for
testing which we are grateful for. The authors also would thank Christos Malliakas
for providing the NaCl sample and assistance on the measurements on this sample.
SrRietveld was developed as part of the
DANSE (distributed data analysis of neutron scattering experiments) project,
which is funded by the National Science Foundation under grant DMR-0520547.
Use of the Advanced Photon Source was supported by the U.S. Department of
Energy, Office of Science, Office of Basic Energy Sciences, under Contract
No. W-31-109-Eng-38.

\bibliography{billinge-group,abb-billinge-group,everyone}		
\bibliographystyle{iucr}	

\end{document}